\documentclass{article}

\input{psfig} 

\newcommand{\figlab}[1]{\label{fig:#1}}

\newcommand{\figref}[1]{\ref{fig:#1}}
\newcommand{\eqref}[1]{(\ref{eq:#1})}

\newtheorem{theorem}{Theorem}

{\catcode`\@=11
\gdef\setft#1#2#3{%
\def\@oddfoot{
{\setbox0=\hbox{#1}
\setbox1=\hbox{#3}
\ifdim\wd0>\wd1
\dimen0=\wd0
\box0\hfil#2\hfil\hbox to\dimen0{\hfil\hfil\box1}
\else \dimen0=\wd1
\hbox to\dimen0{\box0\hfil }\hfil#2\hfil\box1 \fi
}}} }

\def\complaint#1{}
\def\withcomplaints{
\newcounter{mycomplaints}
\def\complaint##1{\refstepcounter{mycomplaints}%
\ifhmode%
\unskip%
{\dimen1=\baselineskip \divide\dimen1 by 2 %
\raise\dimen1\llap{\tiny -\themycomplaints-}}\fi%
\marginpar{\tiny [\themycomplaints]: ##1}}%
}

\usepackage{latexsym} 

\title{Computational Geometry Column 36}
\author{%
Joseph O'Rourke\thanks{
Dept. of Computer Science, Smith Col\-lege, North\-ampton, 
MA 01063, USA.
\-orourke@cs.\-smith\-.edu.
Supported by NSF Grant CCR-9731804.
}
}
\date{}
\begin{document}
\maketitle
\pagestyle{empty}
\thispagestyle{empty}

\begin{abstract}
Two results in ``computational origami'' are illustrated.
\end{abstract}

Computational geometry has recently been applied to solve two
open problems in ``origami mathematics.''

\section{One Cut Suffices}

The first result is remarkable in its generality:
\begin{theorem}
Any planar straight-line drawing may be cut out of one sheet
of paper by a single straight cut, after appropriate 
folding~{\em \cite{ddl-fcp-98,ddl-foscs-99}}.
\end{theorem}
The drawing need not be connected; it may include adjoining
polygons, nested polygons, floating lines segments, and isolated points.
The algorithm of Demaine, Demaine, and Lubiw computes
a crease pattern whose folding produces a flat
origami that aligns all edges of
the drawing on the same line $L$.  
Removal of $L$ from the paper ``cuts out'' the drawing.\footnote{
	It is possible that $L$ will lie along folds, e.g., when
	the drawing consists of a single line segment.}

We illustrate the results of their algorithm applied to a polygon in
the shape of the letter {\tt H}
in Fig.~\figref{H}(a1).
Fold the sheet first in half along the horizontal bisector of the {\tt H}~(a1),
and then in half again along the vertical bisector~(a2).
Now many of the polygon's edges lie on top of one another.
Next, fold along the diagonal bisector of the right angle illustrated~(a3)
to align the adjacent edges.
Continuing in this manner, after five folds, all edges lie on
the same vertical line,
and cutting along the arrow
shown in Fig.~\figref{H}(a6) removes the {\tt H}.
\begin{figure}[htbp]
\begin{center}
\ \psfig{figure=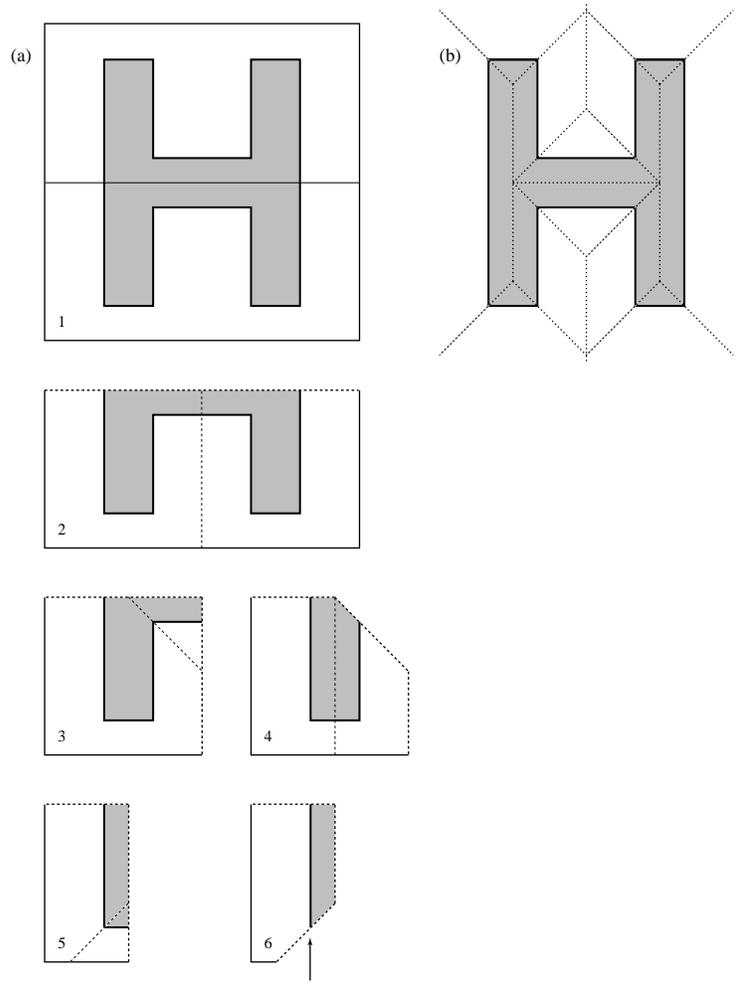,height=13cm}
\end{center}
\caption{(a) Folding to align edges.  The first fold (solid) is a mountain
fold; all others (dashed) are valley folds.
(b) The straight skeleton of the polygon.}
\figlab{H}
\end{figure}

The use of bisectors is a natural technique for overlapping
the edges incident to a vertex, and suggests that the medial axis
or Voronoi diagram may play a role.
In fact the appropriate concept here is the 
{\em straight skeleton}~\cite{aa-ssgpf-96}.
For a polygon, this skeleton is defined by the tracks vertices
follow when the shape shrinks via inward, parallel movement of the
edges.
For the {\tt H}-polygon, the skeleton is particularly simple,
but more complex shapes lead to shrinking 
``events'' which disconnect the shape; then each is shrunk
recursively.
The skeleton may be defined for general straight-line plane graphs,
developing both interior and exterior to faces.
Fig.~\figref{H}(b) shows the complete skeleton for the {\tt H}-shape.

For this simple shape, all creases lie on lines containing skeleton edges.
More complex shapes, for example the butterfly in
Fig.~\figref{butterfly}, require in addition {\em perpendiculars\/}
incident to skeleton vertices, which (perhaps) recursively 
generate more perpendiculars based on other cut edges.
This recursive phenomena means that the number of creases is
unbounded in terms of the number of vertices or 
minimum ``feature'' size of the drawing.
This flaw has subsequently been circumvented by an algorithm
based on disk-packing~\cite{bddh-dpaomt-98}.
\begin{figure}[htbp]
\begin{center}
\ \psfig{figure=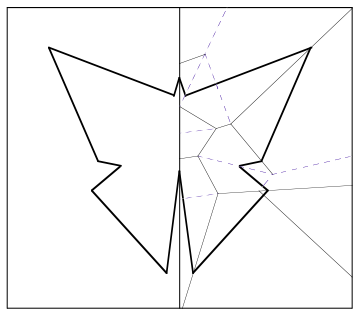,height=4cm}
\end{center}
\caption{Crease pattern to cut out a butterfly. 
[Drawing courtesy of Erik Demaine.]}
\figlab{butterfly}
\end{figure}

\section{Wrapping Polyhedra}

Akiyama posed the question of whether any (perhaps nonconvex) polyhedron
may be ``wrapped'' by a single piece of paper~\cite{a-wtcdg-97}.
Portions of paper may be hidden by folding and tucking under, but
the final result should exactly cover the faces of the polyhedron
without requiring the paper to be cut.
His question was answered and extended with this theorem:

\begin{theorem}
Any polyhedron may be wrapped with a sufficiently large square sheet
of paper.  This implies that any connected, planar, polygonal region may be
covered by a flat origami folded from a single square of paper.
Moreover, any $2$-coloring of the faces may be realized with paper
whose two sides are those colors~{\em \cite{ddm-ffswpp-99}}.
\end{theorem}
Demaine, Demaine, and Mitchell provide three distinct algorithms
for achieving such a folding, each with different properties and
tradeoffs among desirable quantities.  The first is based on
Hamiltonian triangulations, the second on straight skeletons,
and the third on convex decompositions.
I will illustrate the general idea by folding a silhouette
for a polygon in the shape of the letter {\tt I}.

All three methods share the same first step: 
accordian-fold the paper into a strip; see Fig.~\figref{I}(a).
The methods differ on how this strip is used to cover the
faces.  The convex decomposition method starts with a partition
of the faces into convex pieces, and then covers each face
in the order determined by a traversal of a spanning tree of 
the partition dual.
An optimized version of their algorithm could achieve the
simple covering shown in Fig.~\figref{I}(b).
In this example no particular coloring was sought, but one can
see there is freedom on the choice between mountain and valley
folds, freedom which ultimately can be exploited to
achieve any given $2$-coloring.
\begin{figure}[htbp]
\begin{center}
\ \psfig{figure=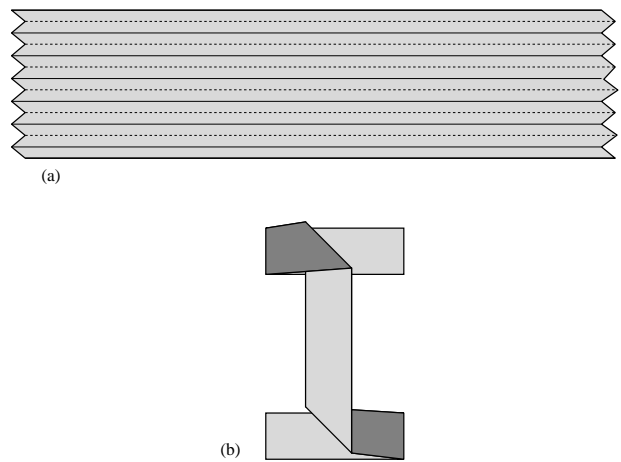,height=6cm}
\end{center}
\caption{Folding a square to cover the shape {\tt I}:
(a) accordian-fold to strip; (b) strip folding.}
\figlab{I}
\end{figure}

Although one method in~\cite{ddm-ffswpp-99}, the ``zig-zag milling''
method that follows a Hamiltonian triangulation, hides an arbitrarily
small fraction of the strip's area, the accordian-fold step wastes
much of the original square paper.  It remains open to achieve the
same universality with a more efficient wrapping.
\bibliographystyle{alpha}
\bibliography{/home1/orourke/bib/geom/geom}

\begin{thebibliography}{BDDH98}

\bibitem[AA96]{aa-ssgpf-96}
O.~Aicholzer and F.~Aurenhammer.
\newblock Straight skeletons for general polygonal figures in the plane.
\newblock In {\em Proc. 2nd International Computing and Combinatorics:
  COCOON'96}, volume 1090 of {\em Lecture Notes Comput. Sci.}, pages 117--126.
  Springer-Verlag, 1996.

\bibitem[Aki97]{a-wtcdg-97}
J.~Akiyama.
\newblock Why {Taro} can do geometry.
\newblock In {\em Proc. 9th Canad. Conf. Comput. Geom.}, page 112, 1997.

\bibitem[BDEH98]{bddh-dpaomt-98}
M.~Bern, E.~D. Demaine, D.~Eppstein, and B.~Hayes.
\newblock A disk-packing algorithm for an origami magic trick.
\newblock In {\em Proc. Internat. Conf. Fun with Algorithms}, Elba, Italy, June
  1998.

\bibitem[DDL98]{ddl-fcp-98}
E.~D. Demaine, M.~L. Demaine, and A.~Lubiw.
\newblock Folding and cutting paper.
\newblock In {\em Proc. Japan Conf. Discrete Comput. Geom.}, Lecture Notes in
  Comput. Sci., Tokyo, Japan, December 1998. Springer-Verlag.

\bibitem[DDL99]{ddl-foscs-99}
E.~D. Demaine, M.~L. Demaine, and A.~Lubiw.
\newblock Folding and one straight cut suffice.
\newblock In {\em Proc. 10th Annu. ACM-SIAM Sympos. Discrete Alg. (SODA'99)},
  pages 891--892, Baltimore, Maryland, January 1999.

\bibitem[DDM99]{ddm-ffswpp-99}
E.~D. Demaine, M.~L. Demaine, and J.~S.~B. Mitchell.
\newblock Folding flat silhouettes and wrapping polyhedral packages: New
  results in computational origami.
\newblock In {\em Proc. 15th Annu. ACM Sympos. Comput. Geom.}, pages 105--114,
  Miami Beach, Florida, June 1999.

\end{thebibliography}
\end{document}